\documentclass[prd,aps,12pt]{revtex4}
\usepackage{latexsym}
\usepackage{amssymb}
\usepackage[dvips]{graphicx}
\usepackage{color}
\usepackage{graphicx}
\begin{document}
\newcommand{\beq}{\begin{equation}}
\newcommand{\eeq}{\end{equation}}
\newcommand{\ie}{{\sl i.e\/}}
\newcommand{\half}{\frac 1 2}
\newcommand{\lag}{\cal L}
\newcommand{\ove}{\overline}
\newcommand{\et}{{\em et al}}
\newcommand{\Prd}{Phys. Rev D}
\newcommand{\Prl}{Phys. Rev. Lett.}
\newcommand{\Plb}{Phys. Lett. B}
\newcommand{\Cqg}{Class. Quantum Grav.}
\newcommand{\Grg}{Grav....}
\newcommand{\Np}{Nuc. Phys.}
\newcommand{\Fp}{Found. Phys.}
\renewcommand{\baselinestretch}{1.2}

\title{The cosmological origins of nonlinear Electrodynamics}

\author{M. Novello and C.E.L. Ducap}
\affiliation{Centro de Estudos Avan\c{c}ados de Cosmologia (CEAC/CBPF) \\
 Rua Dr. Xavier Sigaud, 150, CEP 22290-180, Rio de Janeiro, Brazil}

\date{\today}

\begin{abstract}
We present a mechanism that allows to describe any nonlinear theory of Electrodynamics as a consequence of the coupling of the electromagnetic field to gravity in the presence of a vacuum represented by the cosmological constant. We emphasize gravity\rq s exclusive role of catalysis.
 \end{abstract}

\vskip2pc
 \maketitle

\section{Introduction}

 Recently \cite{novello} \cite{novello3} one of us has examined how a specific self-interaction of a fermionic field, in the Heisenberg description of a dynamical model of elementary particles, can be related to cosmology. To be more specific, it was proved that nonlinearity of Heisenberg \rq s dynamics  can be interpreted as a consequence of the existence of a cosmological constant $ \Lambda$ through the  influence of a gravitational process. Such method has also been the main tool to display a gravitational mechanism to describe the origin of the masses of all particles as shown in \cite{novelloM1} \cite{novelloM2}.

 The presence of $ \Lambda$ is to be treated, according to Einstein\rq s general relativity as a global property of the universe. In modern framework  $ \Lambda$ can be alternatively interpreted as some sort of influence of the vacuum configuration. For our purposes here either one of these interpretations is allowed, once the mechanism which we will present does not distinguishes between either interpretation. Nevertheless, just to be specific we will follow Einstein-Mach and interpret $\Lambda$ as the way the rest-of-the-universe affects any material body or energy of any kind.

In the present work we apply this same idea to analyze the effects of $ \Lambda$ on the electromagnetic field. We will show that any non linear theory of Electrodynamics can be interpreted as a consequence of the gravitational interaction of the electromagnetic field $ F_{\mu\nu}$ with the rest-of-the-universe. Let us point out from the very beginning that the function of gravity in this process is just to act as a catalist, that is an intermediary agent between the cosmical vacuum and the electromagnetic field. Our starting point concerns the generalization of Mach\rq s principle \cite{dicke} and state that not only inertia depends on the homogeneous distribution of the energy-momentum throughout all space, but others properties of bodies of any nature and any distribution of energy may also be influenced by $\Lambda.$

 We follow here a similar procedure and use such Extended Mach Principle \cite{novello3} \cite{bouncing} to analyze the action of the cosmic energy-momentum distribution represented by
$$
T_{\mu\nu} = \Lambda \, g_{\mu\nu}
$$
and its induction of nonlinear processes. In the present paper we consider the case of nonlinear Electrodynamics.

 In short we shall prove that non-minimal coupling $ F_{\mu\nu}$ to gravity in the realm of general relativity generates all forms of self-interaction of the electromagnetic field. We shall see that gravity acts only to promote a catalysis that allows the interaction of electromagnetic field and the rest-of-the-universe through $ \Lambda.$

\section{The scenario}

The minimal coupling principle states that in the framework of General Relativity the dynamics of the metric and of Maxwell theory of Electrodynamics is provided by the action

\begin{equation}
S = \int \, \sqrt{- g} \left( - \, \frac{c^{2}}{4} \,F + \frac{1}{2 \, \kappa}\, R \,\right) d^{4} x
\label{1}
\end{equation}
where
$$ F \equiv F_{\mu\nu} \, F^{\mu\nu}. $$
In the case of non minimal coupling we set the general form of the Lagrangian as given by
\begin{equation}
L = - \, \frac{c^{2}}{4} \,F + \frac{1}{2 \,\kappa} \, R + \frac{1}{\kappa} \, \Lambda  + \frac{1}{\kappa} \, V(F) \, R + L_{CT}
\label{2}
\end{equation}
from now on we set $ c = 1$ for the velocity of light. Note that $ V(F) $ is dimensionless and represents the non-minimal coupling with gravity; the counter-term $L_{CT}, $ which depends only on $ F $ and its first derivatives,  is introduced in order to eliminate in the equation of motion of the electromagnetic field the presence of higher order derivatives.
The most general form of this counter-term is

\begin{equation}
L_{CT} = H(F) \, \partial_{\mu} F \, \partial^{\mu} F
\label{3}\end{equation}

In order to accomplish this task and to avoid terms of second order derivatives in the dynamics of $ F^{\mu\nu}$ the function $ H $ must depend on $ V$
and if we set $ V = 0 $ then $ H $ vanishes. This dynamics represents the electromagnetic field coupled non-minimally with
gravity. Although at the very beginning it seems that there is no direct influence of $ \Lambda$ on the electromagnetic field, we shall see that due to the non-minimal coupling such an influence will appear and, more than this, it will be the real responsible for the rising of nonlinearities of the electromagnetic field, even when we take the limit of vanishing gravity. We limit our analysis here to the case in which the nonlinearity appears only on functions of the invariant $ F.$

Independent variation of the electromagnetic potential $ A_{\mu} $ and $g_{\mu\nu}$ yields
\begin{equation}
F^{\mu\nu}{}_{; \nu} - \frac{4}{\kappa} [ R \, V^{'} \, F^{\mu\nu}]_{; \nu} -4 [ H^{'} \, F_{\lambda} \, F^{\lambda} \, F^{\mu\nu}]_{; \nu} + 8 [ (H \, F^{\lambda})_{; \lambda} \, F^{\mu\nu}]_{; \nu} = 0
 \label{4}
\end{equation}
where $ V^{'} = dV/dF$ represents the derivative with respect to $F.$ For the metric we obtain the equation

\begin{eqnarray}
&&\frac{1}{\kappa} \, (1 + 2 \, V) \,( R_{\mu\nu} - \frac{1}{2} \, R \, g_{\mu\nu} ) =
-  \,E_{\mu\nu}  + \frac{\Lambda}{\kappa} \, g_{\mu\nu} \nonumber \\
&-& (\frac{4}{\kappa} \, R \,V^{'} + 4 \, H^{'} \, F_{\alpha} \, F^{\alpha} - 8 \, H \, \Box F)  \, F_{\mu\alpha} \, F_{\nu}{}^{\alpha} \nonumber \\
&+& \frac{2}{\kappa} \,\Box  V \, g_{\mu\nu} - \frac{2}{\kappa} \, V_{,\mu; \nu} - 2 H \, F_{\mu}  \, F_{\nu}   \label{6}
\end{eqnarray}

where
$$ E_{\mu\nu} = F_{\mu\alpha} \, F^{\alpha}{}_{\nu} + \frac{1}{4} \, F \, g_{\mu\nu} $$
and $ F_{\mu} = \partial_{\mu} F. $ Once $ V = V(F)$ it follows $ \Box V = V^{'} \, \Box F + V^{''} \, F_{\alpha} \, F^{\alpha}.$  The trace of equation (\ref{6}) yields

\begin{eqnarray}
&&R \, (1 + 2 \, V - 4 \, F \, V^{'}) = - \, 4 \, \Lambda
-  \Box F ( 6 \, V^{'} +  8 \, H \, F) \nonumber \\
&-& F_{\mu} \, F^{\mu} ( 6 \, V^{''} +2 \, \kappa \, H + 4 \, \kappa \,H^{'} \, F)
\label{7}
\end{eqnarray}
or,

\begin{equation}
R = \frac{- \, 1}{(1 + 2 \, V - 4 \, F \, V^{'})} \left( 4 \, \Lambda +  \Box F ( 6 \, V^{'} +  8 \, H \, F)   + F_{\mu} \, F^{\mu} ( 6 \, V^{''} +2 \, \kappa \, H + 4 \, \kappa \,H^{'} \, F) \right)
\label{8}
\end{equation}
Using this result in the equation of the electromagnetic field (\ref{4}) gives

\begin{equation}
X \, F^{\mu\nu}{}_{; \nu} + \Omega \, F^{\mu\nu} \, F_{\nu} = 0.
\label{9}
\end{equation}
where
$$ X = 1 - \frac{4}{\kappa} \, R \, V^{'} + 4 \, H^{'}\, F^{\alpha} \, F _{\alpha} + 8 \, H \, \Box F$$

Using the equation of the trace of the scalar of curvature we obtain

\begin{eqnarray}
 X &\equiv& 1 +\frac{16 \, \Lambda \,V^{'} }{\kappa\, (1 + 2 \, V- 4 \, F \, V^{'})}   \nonumber \\
&+&  F_{\alpha} \, F^{\alpha} \, \left( 4 \, H^{'}  + \frac{24 \, V^{'} \,V^{''} + 8 \,\kappa \, H \, V^{'} + 16 \kappa \, F \, H^{'} \, V^{'}}{(1 + 2 \, V- 4 \, F \, V^{'})}\right) \nonumber \\
 &+& \Box F \, \left( 8 \, H + \frac{24 \, (V^{'})^{2} + 32 \, F \, H}{(1 + 2 \, V) - 4 \, F \, V^{'}} \right)
\label{10}
\end{eqnarray}
 and
\begin{equation}
\Omega \, F_{\nu}= - \frac{4}{\kappa} \, ( R \, V^{'})_{;\nu} + 4 \, (H^{'}\, F_{\lambda}\, F^{\lambda})_{; \nu} + 8 \, (H \, \Box F)_{;\nu}
\label{11}
\end{equation}

At this stage it is worth to select among all possible candidates of
$ V$ and $ H $ particular ones that makes the factor on the gradient of $ F $ and $ \Box  F $ to disappear from the expressions of $ X $  and $ \Omega.$ This provides four conditions involving functions $ V $ and $ H.$ However a direct calculation shows that these four equations are related and provides only one condition relating the dependence of $ V $ and $ H.$ Indeed, the factor on $ X $ multiplying $ \Box F$ gives

\begin{equation}
H + 3 \, \frac{(V^{'})^{2}}{\kappa \,(1 +  2\, V)}= 0.
\label{12}
\end{equation}
and the other condition on $ X $ that is the factor multiplying $  F_{\mu} \, F^{\mu} $ yields

\begin{equation}
H^{'} +  \frac{2 \,H \, V^{'}}{(1 +  2\, V)} + \frac{6 \, V^{'} \, V^{''}}{\kappa \,(1 +  2\, V)}= 0.
\label{13}
\end{equation}

By a simple inspection we recognize that the second equation is nothing but the derivative of equation (\ref{12}). Then $ X $ reduces to

\begin{equation}
X = 1 + \frac{16 \, \Lambda}{\kappa} \, \left(\frac{V^{'}}{1 +  2\, V - 4 \, F \, V^{'}} \right).
\label{2482016}
\end{equation}

The beautiful result comes next: the above condition (\ref{12}) eliminates also all unpleasant terms containing gradients and  $ \Box F $ appearing on $ \Omega.$  Indeed, we have

$$ \Omega =   \frac{d}{d F} X.$$
which allow us to re-write equation (\ref{4}) for the electromagnetic field under the form

\begin{equation}
L_{F} \, F^{\mu\nu}_{; \nu} + L_{F F} \, F^{\mu\nu} \, F_{\nu} = 0.
\label{14}
\end{equation}
where $ L_{F} = dL/ dF \equiv L^{'}.$

Thus as a result of the gravitational interaction the electromagnetic field obeys a non-linear equation of motion characterized by the Lagrangian $ L (F).$ Indeed from the identification
$$  X =  L_{F}$$
into equation (\ref{9}) and using the connection between $ V$ and $ H $ given by equation (\ref{12}) it follows that the non-linear Electrodynamics can be obtained by the integration of the expression
\begin{equation}
\frac{ 4 \, V^{'}}{ 1 + 2 \, V} = \frac{ L^{'} - 1}{F \, (L^{'} - 1) + 4 \, \Lambda /\kappa}.
\label{15}
\end{equation}
Setting
$$ Z(F) \equiv \frac{ L^{'} - 1}{F \, (L^{'} - 1) + 4 \, \Lambda /\kappa} $$
the form of $ V $ is given by

\begin{equation}
2 \, V = \exp \Delta - 1.
\label{2429}
\end{equation}
where
$$ \Delta = \frac{1}{2} \, \int Z dF.$$

Thus given any non-linear Lagrangian $ L(F)$ we determine the values of $ V $  and $ H $ through which the non-minimal interaction with gravity induces such nonlinearities. In the other way round, specification of the dependence of $ V(F)$ provides the form of the equivalent non-linear Lagrangian.

Let us point out that this is possible due to the influence of the rest-of-the-Universe represented by the cosmological constant on the electromagnetic field. If $ \Lambda $ vanishes then the self-interaction disappears and we are left with linear Maxwell \rq s dynamics. Indeed from eq. (\ref{2482016}) when $ \Lambda = 0 $ then $ X = 1$ yielding the linear Maxwell dynamics.

\section{Final remarks}

In this work we analyzed the influence of all the material
content of the universe on the electromagnetic field described by a homogeneous energy distribution $ T_{\mu\nu} = \Lambda g_{\mu\nu}.$  When
$\Lambda $ vanishes,  the dynamics of the field is independent of
the global properties of the universe and it reduces to the Maxwell \rq s equations. In the second case, the rest-of-the-universe induces a non-linear dynamics yielding the equation of motion (\ref{14}).

Such scenario is a particular example of generalization of Mach \rq s principle setting a global mechanism by means of which the influence of the global structure of the universe appears in any physical theory . We would like to emphasize gravity\rq s catalysis role in the generation of nonlinearities for the electromagnetic field.

Although we have analyzed the case in which the electromagnetic Lagrangian depends only on the invariant $ F $ the generalization for the case in which one introduces the dependence on $ G = F_{\mu\nu}^{*} \, F^{\mu\nu} $ follows along the same lines. Finally let us remind that the term rest-of-the-universe that we used, concerns the environment of any body $ \mathbb{A}, $ that is the whole domain of influence on $ \mathbb{A} $ of the remaining bodies in the universe. Note that it is completely irrelevant -- for the gravitational mechanism employed in this paper --- if parameter $\Lambda$ has a classical global origin identified with the cosmological constant introduced by Einstein or
a local quantum one identified with the vacuum of quantum fields.

\section*{Acknowledgements}
 This work was partially supported by {\em Conselho Nacional de Desenvolvimento
Cient\'{\i}fico e Tecnol\'ogico} (CNPq) and \emph{Coordena\c{c}\~ao do Aperfei\c{c}oamento do Pessoal do Ensino Superior (CAPES)}.


\begin{thebibliography}{100}
\bibitem{novello} M.Novello in Cosmology and Gravitation, Procceedings of the XIV Brazilian School of Cosmology and Gravitation, Eds M. Novello and S E P Bergliaffa, Cambridge Scientific Publishers (2011).
\bibitem{novello3} M.Novello: \emph{The cosmological origin of the Nambu-Jona-Lasinio model} in Int. J. Mod. Phys. A 26 (2011) 3781.
\bibitem{novelloM1} M.Novello: \emph{What is the origin of the mass of the Higgs boson? }in Phys Rev D 86, 063510 (2012)
\bibitem{novelloM2}M.Novello: \emph{The gravitational mechanism to generate mass}, in Class. Quantum Grav. 28 (2011) 035003
\bibitem{dicke} R. H. Dicke: \emph{Mach\rq s principle and a relativistic
theory of gravity} in Relativity, groups and topology (Les Houches,
1964) Gordon and Breach Publishers.
Reports vol 463, n 4, July 2008 and referennces therein.
\bibitem{bouncing} M. Novello and S E P Bergliaffa in Physics Reports vol 463, no 4 (2008) 129.
 \bibitem{heisenberg} W. Heisenberg, Z. Naturforsch. 14, 441 (1959)
and earlier papers quoted there.
\end{thebibliography}
 \end{document}